\documentclass[aps,preprint,epsf,preprintnumbers,nofootinbib,groupedaddress,superscriptaddress,amsmath,amssymb]{revtex4}

\usepackage{dcolumn}% Align table columns on decimal point
\usepackage{bbm}
\usepackage{bm}% bold math
\usepackage{amssymb}
\usepackage{amsmath}
\usepackage{color}
\usepackage{slashed}
\usepackage{hhline}
\usepackage{epsfig}    
\def\be{\begin{equation}}
\def\ee{\end{equation}}
\newcommand{\bea}{\begin{eqnarray}}
\newcommand{\eea}{\end{eqnarray}}
\newcommand{\nn}{\nonumber}

\numberwithin{equation}{section}

\begin{document}

%%%%%%%%%
\title{Dark Matter, Muon $g-2$, Electric Dipole Moments and $Z\to \ell_i^+ \ell_j^-$ in a One-Loop Induced Neutrino Model}

\author{Cheng-Wei Chiang}
\email{chengwei@phys.ntu.edu.tw}
\affiliation{Department of Physics, National Taiwan University, Taipei 10617, Taiwan}
\affiliation{Institute of Physics, Academia Sinica, Taipei 11529, Taiwan}
\affiliation{Physics Division, National Center for Theoretical Sciences, Hsinchu 30013, Taiwan}

\author{Hiroshi Okada}
\email{macokada3hiroshi@cts.nthu.edu.tw}
\affiliation{Physics Division, National Center for Theoretical Sciences, Hsinchu 30013, Taiwan}

\author{Eibun Senaha}
\email{eibunsenaha@ntu.edu.tw}
\affiliation{Department of Physics, National Taiwan University, Taipei 10617, Taiwan}

\date{\today}
\preprint{NCTS-PH/1705}

\begin{abstract}
We study a simple one-loop induced neutrino mass model that contains both bosonic and fermionic dark matter candidates and has the capacity to explain the muon anomalous magnetic moment anomaly.  
We perform a comprehensive analysis by taking into account the relevant constraints of charged lepton flavor violation, electric dipole moments, and neutrino oscillation data.  We examine the constraints from lepton flavor-changing $Z$ boson decays at one-loop level, particularly when the involved couplings contribute to the muon $g-2$.  It is found that $\text{BR}(Z\to \mu\tau)\simeq (10^{-7}$ - $10^{-6})$ while $\text{BR}(\tau\to\mu\gamma)\lesssim 10^{-11}$ in the fermionic dark matter scenario.  The former can be probed by the precision measurement of the $Z$ boson at future lepton colliders.
\end{abstract}
\maketitle
\newpage

%%%%%%%%%%%%%%%%%%%%%%%%%%%%%%%%%%%%%%%%%%%%%%%%%%
\section{Introduction}
%%%%%%%%%%%%%%%%%%%%%%%%%%%%%%%%%%%%%%%%%%%%%%%%%%

Even though the standard model (SM) of particle physics has been very successfully predicting or explaining most experimental results and phenomena, it still has a few outstanding problems with empirical observations.  One is the origin of neutrino mass as inferred from neutrino oscillation experiments.  We still do not know whether it is of Dirac or Majorana type and whether it has a normal or inverted hierarchy.  Another one is the existence of dark matter in the Universe.  Despite its immense gravitational effects at the cosmological scale, we do not know what kind of object it is and how it interacts with SM particles otherwise.

Radiative seesaw models provide one of the promising scenarios to explain the neutrino oscillation data and dark matter candidates simultaneously.
In particular, one-loop models have various unique applications to elementary particle phenomenology such as flavor predictive models~\footnote{For example, some specific two-zero textures are obtained if an additional symmetry is imposed, with the ability to predict the mass hierarchy of neutrinos, the scale of third neutrino mass, and so on~\cite{Fritzsch:2011qv}. Note that here this property does not appear in any two-loop or higher-loop induced radiative neutrino models.} in the neutrino sector~\cite{Baek:2015mna, Ko:2017yrd, Lee:2017ekw} and leptogensis~\cite{Gu:2007ug, Lu:2016ucn, Gu:2010ye}.
However, leptogenesis in this framework tends to rely on the mechanism of resonant effect or a large hierarchy among the neutrino Yukawa couplings in order to satisfy the neutrino oscillation data and evade the washout problem simultaneously.
This could be resolved by introducing extra neutral fermions in such a way that these fermions decouple from the physical spectrum at the electroweak (EW) scale.

In this work, we add to the SM weak isospin singlet and doublet colorless fermions.  We show how to make the six heavy neutral fermions decouple at the EW scale by invoking a modified Casas-Ibarra parametrization method~\cite{Casas:2001sr}, leaving the three light fermions as the active neutrinos.  Furthermore, we discuss how the model can explain the discrepancy of the muon anomalous magnetic moment from the SM prediction and contribute to the relic density of dark matter (DM).  We present a comprehensive analysis to find the parameter space that can satisfy constraints such as charged lepton flavor-violating decays, electric dipole moments, direct detection searches of DM, $Z\to \bar\ell_i\ell_j$ decays, and neutrino oscillation data.
One of the most important predictions of this model is that $\text{BR}(Z\to\mu\tau)$
can be as large as $\mathcal{O}(10^{-6})$,
which is just one order smaller than the current experimental upper bound, while
$\text{BR}(\tau\to\mu\gamma)\lesssim 10^{-11}$ in the fermionic DM scenario. 
Such a large $\text{BR}(Z\to\mu\tau)$ can be probed by precision measurements of the $Z$ boson
at future lepton colliders
such as the ILC~\cite{Baer:2013cma}, FCC-ee~\cite{Gomez-Ceballos:2013zzn},
CEPC~\cite{CEPC-SPPCStudyGroup:2015csa}, etc.

This paper is organized as follows.
Sec.~\ref{sec:model} introduces our model and gives the relevant formulas of various physical quantities.
Sec.~\ref{sec:NA} presents a comprehensive analysis that takes into account many current data to constrain the parameter space of the model.  We have separate discussions about bosonic and fermionic DM candidates in the model.
We summarize our conclusions in Sec.~\ref{sec:summary}.

%%%%%%%%%%%%%%%%%%%%%%%%%%%%%%%%%%%%%%%%%%%%%%%%%%
\section{Model setup \label{sec:model}}
%%%%%%%%%%%%%%%%%%%%%%%%%%%%%%%%%%%%%%%%%%%%%%%%%%

%%%%%%%%%%
\begin{widetext}
\begin{center} 
\begin{table}[tb]
\begin{tabular}{|c||c|c|c|c||c|c|c|}\hline\hline  
&\multicolumn{4}{c||}{Lepton Fields} & \multicolumn{3}{c|}{Scalar Fields} \\\hline
& ~$L_L$~ & ~$e_R^{}$~ & ~$L'$ ~ & ~$N_R$~ & ~$\Phi$~  & ~$S$~ & ~$\varphi$ \\\hline 
$SU(2)_L$ & $\bm{2}$  & $\bm{1}$ & $\bm{2}$ & $\bm{1}$ & $\bm{2}$ & $\bm{1}$ & $\bm{1}$ \\\hline 
$U(1)_Y$ & $-1/2$ & $-1$  & $-1/2$ & $0$& $1/2$ & $0$ & $0$  \\\hline
$Z_2$ & $+$ & $+$   & $-$ & $-$& $+$& $-$  & $+$  \\\hline
\end{tabular}
\caption{Contents of colorless fermion and scalar fields in the model,
and their charge assignments under $SU(2)_L\times U(1)_Y\times Z_2$.}
\label{tab:1}
\end{table}
\end{center}
\end{widetext}
%%%%%%%%%%

In this section, we describe the setup of our model.  We only introduce new colorless fields to the SM.
The contents of fields without participating in the strong interactions and their charges are given in Table~\ref{tab:1}.
We add three isospin-doublet, vector-like exotic fermions $L'$, three isospin-singlet, Majorana fermions $N_R$, and two isospin-singlet scalars $S$ and $\varphi$ to the SM.~\footnote{In fact, the minimal setup is to have only two species for each of $L'$ and $N_R$ if the lightest neutrino is massless.}  Here $S$ has to be a complex field to induce nonzero neutrino mass, but $\varphi$ can be a real field for simplicity.
%%% 
We assume that only the SM-like Higgs field $\Phi$ and the new real scalar $\varphi$ develop nonzero vacuum
expectation values (VEV's), denoted by $v/\sqrt2$ and $v'$, respectively. 
We also impose a $Z_2$ symmetry, under which only $L'$, $N_R$, and $S$ have odd parity, to ensure the stability of DM candidate(s).  In the case of a fermionic DM candidate in the model, $\varphi$ plays an important role in explaining its relic density.

The relevant Yukawa Lagrangian under these symmetries is given by
\begin{align}
-\mathcal{L}_{Y}
=&
(y_{\ell})_{ij} (\bar L_{L})_i \Phi (e_{R})_j
+ f_{ij} (\bar L_{L})_i (L'_{R})_j S
+ g_{ij} (\bar L'_{L})_i (N_{R})_j \tilde{\Phi}
+ \frac{(y_{N})_{ij}}2 \varphi (\bar N_{R}^c)_i (N_{R})_j 
\nn\\
&\quad
+ (y_{E})_{ij} \varphi (\bar L'_{L})_i (L'_{R})_j    
+ \frac{ (M_{N})_i } 2 (\bar N_{R}^c)_i (N_{R})_i
+ { (M_{L})_i} (\bar L'_{L})_i (L'_{R})_i 
+ {\rm h.c.} ~, 
\label{Eq:lag-flavor}
\end{align}
where $i, j \in \{1,2,3\}$ are the species indices and $\tilde{\Phi} \equiv i \sigma_2 \Phi^*$, with $\sigma_a ~ (a = 1,2,3)$ denoting the Pauli matrices.  The first term of $\mathcal{L}_{Y}$ generates the SM
charged-lepton masses $m_\ell = y_\ell v/\sqrt2$ after the EW spontaneous breaking of $\Phi$.
Notice that here $M_N$ and $M_L$ are assumed to be diagonal from the beginning without loss of generality.

In the following, we divide our discussions into subsections on the scalar potential, the exotic fermion sector, the neutrino mass, flavor-violating radiative lepton decays, the muon anomalous magnetic moment and electric dipole moments, the $\gamma\gamma$ decay mode of the Higgs boson, flavor-changing leptonic $Z$ decays, and the dark matter candidates.

%%%%%%%%%%%%%%%%%%%%%%%%%%%%%%%%%%%%%%%%%%%%%%%%%%
\subsection{Scalar Potential}
%%%%%%%%%%%%%%%%%%%%%%%%%%%%%%%%%%%%%%%%%%%%%%%%%%

The most general gauge-invariant scalar potential at a renormalizable level is
\begin{align}
{\cal V}(\Phi, \varphi, S)&=
m^2_{\Phi}|\Phi|^2+ \frac{\lambda_{\Phi}}{4} |\Phi|^4
+\mu_\varphi^3 \varphi+ \frac{m^2_\varphi}2 \varphi^2
+  \frac{\mu_2} {3}\varphi^3 
 + \frac{\lambda_{\varphi}}{4}\varphi^4+ {m^2_{S_2}} |S|^2 +\lambda_S|S|^4 \nn\\
&\quad 
+\mu_{\Phi\varphi}|\Phi|^2 \varphi +\frac{\lambda_{\Phi\varphi}}{2} |\Phi|^2\varphi^2
+{\lambda_{\Phi S}} |{\Phi}|^2 |S|^2
+  {\mu_{S_2}} |S|^2\varphi 
+\frac {\lambda_{S\varphi_2}}2 |S|^2 \varphi^2 \nn\\
&\quad + 
\left(\frac{m^2_{S_1}}2 S^2+ \frac{\mu_{S_1}}2 S^2\varphi +\frac{\lambda_{S_1}}{4} S^4
+\frac{\lambda_{S_2}}{3} |S|^2S^2 + \frac{\lambda'_{\Phi S}}2 |{\Phi}|^2 S^2
+\frac{\lambda_{S\varphi_1}}4 S^2 \varphi^2 +{\rm h.c.}\right),
\label{Vtree}
\end{align}
where the scalar fields can be parameterized as 
\begin{align}
&\Phi =\left[
\begin{array}{c}
w^+\\
\frac{v+h+iz}{\sqrt2}
\end{array}\right],\quad 
\varphi=v'+\sigma,
\quad S=\frac{S_R+i S_I}{\sqrt{2}},
\label{component}
\end{align}
where $v\simeq 246$~GeV is the VEV of the Higgs doublet, and $w^\pm$ and $z$ are respectively the Nambu-Goldstone (NG) bosons that become the longitudinal components of $W$ and $Z$ bosons after the EW symmetry breaking.
For the SU(2)-singlet fields, $\varphi$ is assumed to develop the VEV $v'$, while $S$ is inert to be consistent with the $Z_2$ symmetry.

The terms in the last line of Eq.~\eqref{Vtree} yield a mass splitting between $S_R$ and $S_I$.
In this analysis, we assume that $m_{S_1}^2 \neq 0$ and $\mu_{S_1} = \lambda_{S_1} = \lambda_{S_2} = \lambda'_{\Phi S} = \lambda_{S\varphi_1} = 0$ for simplicity.
Therefore, the masses of $S_R$ and $S_I$ are respectively reduced to
\begin{align}
\begin{split}
m_{S_R}^2 & = m_{S_2}^2+m_{S_1}^2+\frac{\lambda_{\Phi S}}{2}v^2
	+\frac{\lambda_{S\varphi_2}}{2}v'^2
+\mu_{S_2}v' ~,
\\
m_{S_I}^2 & = m_{S_2}^2-m_{S_1}^2+\frac{\lambda_{\Phi S}}{2}v^2
	+\frac{\lambda_{S\varphi_2}}{2}v'^2+\mu_{S_2}v' ~.
\end{split}
\end{align}

Imposing the tadpole conditions: $\partial\mathcal{V}/\partial h|_{v}=0$ and $\partial\mathcal{V}/\partial\sigma|_{v'}=0$,
the resulting mass eigenvalues and mixing matrix for the CP-even boson mass matrix
\begin{align}
M_H(h,\sigma)
= \begin{bmatrix}
m^2_{hh} & m^2_{h\sigma} 
\\  
m^2_{h\sigma} & m^2_{\sigma\sigma}
\end{bmatrix}
\end{align}
are respectively given by~\cite{Chiang:2015fta} 
 \begin{align}
&O^T(\alpha)M_H(h,\phi) O(\alpha) 
= \begin{bmatrix} m^2_{H_1} & 0 \\ 0 & m^2_{H_2} \end{bmatrix} 
~,
\end{align}
with
\begin{align}
O 
=
\begin{bmatrix} 
\cos\alpha & -\sin\alpha 
\\ \sin\alpha & \cos\alpha 
\end{bmatrix}
\label{Omix}
~\mbox{and}~
\sin2\alpha = \frac{2 m^2_{h\sigma}}{ m^2_{hh}- m^2_{\sigma\sigma}} ~,
\end{align}
where $H_1$ is the SM-like Higgs ({\it i.e.}, $m_{H_1}=125$~GeV) and $H_2$ is the additional CP-even Higgs boson.  Notice that here $m^2_{hh}$, $m^2_{\sigma\sigma}$, $m^2_{h\sigma}$ as well as $m_{H_i}(i=1,2)$ can be  rewritten in terms of the parameters in the Higgs potential \eqref{Vtree}.
In our analysis, $m_{H_{1,2}}$ and $\alpha$ are fixed by the tree-level relations.
One-loop contributions can be found in Ref.~\cite{Baek:2012uj}.

In the large $v'$ limit, the Higgs boson masses are reduced to
\begin{align}
m_{H_1}^2 
&\simeq 2 \lambda_\Phi v^2 - \frac{\lambda_{\Phi\varphi}^2 v^2}{2\lambda_\varphi}, \quad
m_{H_2}^2 \simeq 2\lambda_\varphi v'^2+\frac{\lambda_{\Phi\varphi}^2v^2}{2\lambda_\varphi}.
\end{align}
As discussed in Ref.~\cite{EliasMiro:2012ay}, 
vacuum metastability of the SM can be cured by the presence of
doublet-singlet mixing since now $\lambda_\Phi>\lambda_\Phi^{\text{SM}}\equiv m_{H_1}^2/(2v^2)\simeq 1/8$.

%%%%%%%%%%%%%%%%%%%%%%%%%%%%%%%%%%%%%%%%%%%%%%%%%%
\subsection{Exotic Fermion Sector}
%%%%%%%%%%%%%%%%%%%%%%%%%%%%%%%%%%%%%%%%%%%%%%%%%%

We define the isospin-doublet exotic fermion fields as:
\begin{align}
L'_{L(R)}\equiv 
\begin{bmatrix}
N'\\
E'^-
\end{bmatrix}_{L(R)} ~.
\end{align}
The $3\times 3$ mass matrix of the charged exotic fermion, denoted by $M_E$, is then given by $M_E = M_L + y_E v'$, which can be cast into the diagonal $M_E^D$ by a bi-unitary transformation, {\it i.e.},
\begin{align} 
M_E^D = (V_C)_L M_E (V_C^\dag)_R ~,
\end{align}
where $(V_C)_{L,R}$ are the rotation matrices for the left-handed and right-handed charged exotic fermions, respectively.
Nonetheless, without loss of generality, we assume here that $(V_C)_{L}=(V_C)_{R} = \mathbbm{1}$, meaning that $M_E$ is already diagonalized, for simplicity in the numerical analyses.

On the other hand, the $9\times 9$ mass matrix for the neutral fermions in the basis of $[N'^C_R,N'_L,N^C_R]$ is given by 
\begin{align}
M=
%[\bar N'^c_L,\bar N'_R]
\begin{bmatrix}
0 & M_E^\dag & 0 \\
M_E^* &  0 & m_{LR}^*\\
0 &  m_{LR}^\dag & M_N^* \\
\end{bmatrix} ~,
\label{eq:neutral-mat}
%%%
\end{align}
where $M_N = M_{N_R} + y_N v'$ and $m_{LR} = gv/\sqrt2$.
%%%
The mass matrix $M$ can be diagonalized by a $9\times9$ unitary mixing matrix $V_N$ as
$M^D = V_N M V_N^T$ and
\begin{align}
\left[\begin{array}{c}
N'^C_R\\ N'_L\\ N^C_R\\
\end{array}\right]
\equiv V_N^T
\left[\begin{array}{c}
\psi^C_{I R}\\ \psi_{J L}\\ \psi^C_{K R}\\
\end{array}\right],
\end{align}
where $\psi^{(C)}_i$ ($i = I, J, K$) are the mass eigenstates, each of which has three components.  In what follows, we will use $\psi_a$ with $a = 1 - 9$ to refer to the nine physical components of neutral fermions.
To obtain an explicit $V_N$ for the numerical analyses, we assume $m_{LR}$ and $M_N$ to be diagonal for simplicity.  With the assumed diagonal $M_E$, $m_{LR}$ and $M_N$, one can diagonalize Eq.~(\ref{eq:neutral-mat}) via a $3 \times 3$ matrix for each ``generation'' of the neutral fermions.

%%%%%%%%%%%%%%%%%%%%%%%%%%%%%%%%%%%%%%%%%%%%%%%%%%
\subsection{Neutrino Mass}
%%%%%%%%%%%%%%%%%%%%%%%%%%%%%%%%%%%%%%%%%%%%%%%%%%

%------------------------------------------------------------------------
\begin{figure}[t]
\centering
\includegraphics[width=8cm]{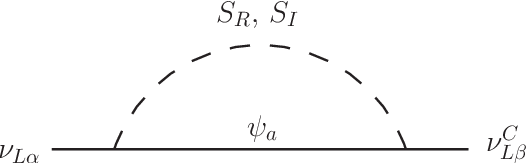}
\caption{One-loop induced Majorana neutrino mass in the model.}
\label{fig:Mnu_1L}
\end{figure}
%------------------------------------------------------------------------

First, we rewrite the terms relevant for the neutrino mass in terms of the mass eigenstates as
\begin{align}
-\mathcal{L}_{Y}
&\ni
{F_{ia}} (\bar \nu_{L})_i P_R \psi_a (S_R + i S_I) 
~~\mbox{with}~~
F_{ia} = \frac{1}{\sqrt2} \sum_{j=1}^{3} f_{ij} (V_N^\dag)_{ja} ~,
\label{Eq:lag-mass}
\end{align}
where $P_R = \frac12 (1 + \gamma_5)$ is the right-handed projection operator.
Then the dominant contribution to the active neutrino mass matrix $m_\nu$ is given at the one-loop level, as shown in Fig.~\ref{fig:Mnu_1L}.  The explicit mass formula is given by 
\begin{align}
&(m_{\nu})_{\alpha\beta}
=\sum_{a=1}^{9}
\frac{F_{\alpha a} M_a F_{\beta a}}{(4\pi)^2}
\left[ \frac{m^2_{S_R}}{m^2_{S_R}-M^2_a} \ln \frac{m^2_{S_R}}{M^2_a}-
\frac{m^2_{S_I}}{m^2_{S_I}-M^2_a} \ln \frac{m^2_{S_I}}{M^2_a} 
 \right] ~,
\end{align}
summing all possible neutral fermions running in the loop.  The structure of this formula is the same as that of a Ma model~\cite{Ma:2006km} except for the rank of the mass matrix $M_N$.

The observed mixing matrix, the Pontecorvo-Maki-Nakagawa-Sakata (PMNS) matrix $U_{\rm PMNS}$~\cite{Maki:1962mu}, can always be realized by introducing the Casas-Ibarra parametrization~\cite{Casas:2001sr}, given by
\begin{align}
(F)_{3\times9}
&=
U_{\rm PMNS}^* 
%%%
\begin{bmatrix}
m_{\nu_1}^{1/2} & 0 &0 \\ 0 & m_{\nu_2}^{1/2} &0 \\ 0 & 0 & m_{\nu_3}^{1/2}
\end{bmatrix}
{\cal O} R ^{-1/2} ~,~
\mbox{or } 
f_{3\times9}= \sqrt2 F V_N ~,
%%%
\nn \\
&\mbox{with }
R_{aa}\equiv \frac{M_a}{(4\pi)^2}
 \left[ \frac{m^2_{S_R}}{m^2_{S_R}-M^2_a} \ln \frac{m^2_{S_R}}{M^2_a}-
\frac{m^2_{S_I}}{m^2_{S_I}-M^2_a} \ln \frac{m^2_{S_I}}{M^2_a}
 \right],
\end{align}
where ${\cal O}$ is a $3\times9$ orthogonal matrix with complex values, which can be decomposed into three $3\times 3$ matrices ${\cal O}\equiv {\cal O}_{\bf 1} +{\cal O}_{\bf 2}+{\cal O}_{\bf 3}$, each of which is orthogonal with complex components as ${\cal O}$.
However, since the last six columns of the mass matrix do not contribute to the active neutrino masses, 
we assume them to have null components; {\it i.e.}, ${\cal O}_{\bf 2}={\cal O}_{\bf 3}={\bf 0}$.
Therefore, we have the parameterization
\begin{align}
{\cal O}_{\bf 1}  =\left[\begin{array}{ccc } {c_{13}}c_{12} &c_{13}s_{12} & s_{13} \\
 -c_{23}s_{12}-s_{23}s_{13}c_{12} & c_{23}c_{12}-s_{23}s_{13}s_{12}& s_{23}c_{13} \\
  s_{23}s_{12}-c_{23}s_{13}c_{12} & -s_{23}c_{12}-c_{23}s_{13}s_{12} & c_{23}c_{13} \\  \end{array}\right],
\end{align}
where $s(c)_{ij}\equiv \sin(\cos)\delta_{ij}$ ($i,j=1,2,3$).
%%%
It also implies that the six heavy neutral fermions can assume any large mass eigenvalues $M_{4-9}$.
%%%
Then the neutrino mass eigenvalues $m_\nu^{{\rm diag.}}\equiv(m_{\nu_1},m_{\nu_2},m_{\nu_3})$ is given by
\begin{align}
m^\dag_\nu m_\nu
&=U_{\rm PMNS}
%%%
\begin{bmatrix}
m_{\nu_1}^{2} & 0 &0 \\ 0 & m_{\nu_2}^{2} &0 \\ 0 & 0 & m_{\nu_3}^{2} \\ 
\end{bmatrix}
U_{\rm PMNS}^\dag ~,
\end{align}
which is subject to the constraints of neutrino oscillation data in Table 1 of Ref.~\cite{Gonzalez-Garcia:2014bfa}:
\begin{align}
\sin^2{\theta_{12}} = 0.304 ~,~ \sin^2{\theta_{23}} = 0.452 ~,~ \sin^2{\theta_{13}}=0.0218 ~,
~\delta_{\text{PMNS}}=\frac{306}{180}\pi.
\end{align}
We take the Majorana CP-violating (CPV) phases to be zero.  Furthermore, in our numerical analysis 
we take the following neutrino masses as an explicit example:
\begin{align}
m_{\nu_1}=0~\text{eV},\quad m_{\nu_2}=\sqrt{0.750}\times 10^{-2}~\text{eV},\quad
m_{\nu_3}=\sqrt{24.57}\times 10^{-2}~\text{eV}.
\end{align}

%%%%%%%%%%%%%%%%%%%%%%%%%%%%%%%%%%%%%%%%%%%%%%%%%%
\subsection{Radiative Lepton Decays with Flavor Violation}
\label{lfv-lu}
%%%%%%%%%%%%%%%%%%%%%%%%%%%%%%%%%%%%%%%%%%%%%%%%%%

%------------------------------------------------------------------------
\begin{figure}[t]
\centering
\includegraphics[width=7cm]{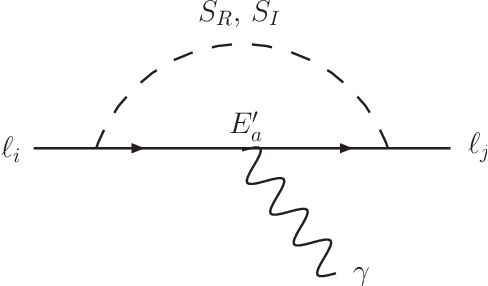}
\caption{LFV processes induced at the one-loop level in the model.}
\label{fig:LFVs}
\end{figure}
%------------------------------------------------------------------------

Lepton flavor-violating (LFV) processes arise from the Yukawa term with the matrix coefficient $f$:
\begin{align}
{\cal L}_Y
\ni
F'_{ia} \bar \ell_{i} P_R {E'_a} (S_R + i S_I) + \mbox{h.c.}
~~\mbox{with}~~
F'_{ia} = \frac{1}{\sqrt2} f_{ij} (V_C^\dag)_{ja} ~,
\label{eq:lvs-g2}
\end{align}
where $(\ell_1,\ell_2,\ell_3)\equiv (e,\mu,\tau)$.
A generic one-loop radiative LFV decay process is plotted in Fig.~\ref{fig:LFVs}.  The corresponding decay branching ratio is given by (for $i \neq j$)
\begin{align}
{\rm BR}(\ell_i\to\ell_j\gamma)
=
\frac{48\pi^3\alpha_{\rm em} C_{ij}}{G_F^2}
\left|\sum_{a=1}^3\sum_{J=R,I} \frac{ F'_{ja} {F'_{i a}}^*}{32\pi^2}
\frac{2+3r_{aJ}-6 r_{aJ}^2 +r_{aJ}^3+6r_{aJ} \ln r_{aJ}}
{6m_{S_J}^2 (1-r_{aJ})^4}
\right|^2 ~,
\label{eq:damu1}
\end{align}
where the fine structure constant $\alpha_{\rm em} \simeq 1/128$, the Fermi constant $G_F \simeq 1.17\times 10^{-5}$ GeV$^{-2}$, $(C_{21},C_{31},C_{32}) \simeq (1,0.1784,0.1736)$, and $r_{aJ}\equiv (M_{E_a}/m_{S_J})^2$.
The current experimental upper bounds at 90\% confidence level (CL) are~\cite{TheMEG:2016wtm, Adam:2013mnn}
\begin{align}
{\rm BR}(\mu\to e\gamma) < 4.2\times10^{-13} ~,~
{\rm BR}(\tau\to e\gamma) < 3.3\times10^{-8} ~,~
{\rm BR}(\tau\to \mu\gamma) < 4.4\times10^{-8} ~.
\end{align}
Note that any constraints on lepton flavor-violating processes $\ell_i \to \ell_j \ell_k \ell_\ell$ at the one-loop level are less stringent than those of $\ell_i \to \ell_j \gamma$ given above~\cite{Toma:2013zsa}. 
{Also, processes such as $\tau \to \mu \nu \bar\nu$ may arise from penguin diagrams by replacing $\gamma$ in Fig.~\ref{fig:LFVs} with the $Z$ boson. However, such deviations will be smaller than the current bounds.  Thus, we do not pursue them hereafter.}

We note in passing that the interaction Eq.~(\ref{eq:lvs-g2}) together with a $H_1$-$S_R$-$S_I$
vertex gives rise to $H_1\to \mu\tau$ at the one-loop level.
In this model, $H_1\to \mu\tau$ mode is proportional to either $m_\mu$ or $m_\tau$
due to the chiral structures of the $\mu$-$E'_a$ and $\tau$-$E'_a$ couplings, 
resulting in $(m_{\mu, \tau}/m_{H_1})^2$ suppressions other than an ordinary 
one-loop suppression factor in this decay. 
It is thus hard to obtain $\text{BR}(H_1\to \mu\tau)\simeq \mathcal{O}(0.1)\%$, which is 
hinted at by the recent LHC data~\cite{Khachatryan:2015kon,Aad:2016blu}.

%%%%%%%%%%%%%%%%%%%%%%%%%%%%%%%%%%%%%%%%%%%%%%%%%%
\subsection{Anomalous Magnetic Moment of Muon and Electric Dipole Moments}
%%%%%%%%%%%%%%%%%%%%%%%%%%%%%%%%%%%%%%%%%%%%%%%%%%

The discrepancy of the muon $g-2$ between the experimental measurement and the SM prediction is given by~\cite{Hagiwara:2011af} 
\begin{align}
\Delta a_{\mu}=(26.1 \pm 8.0)\times 10^{-10}. 
\end{align}
%%%
In our model, the leading contribution comes from the same term in Eq.~(\ref{eq:lvs-g2}) at the one-loop level as discussed in the previous subsection.  Its form is found to be{~\cite{Miller:2007kk, Lindner:2016bgg, Jegerlehner:2009ry}}
 \begin{align}
 \Delta a_\mu^{(1)} \approx 
 \sum_{a=1}^{3} \sum_{J = R,I}
\frac{| F'_{2 a}|^2}{16\pi^2} \int_0^1dx
\frac{x^2(1-x)}{x(x-1)+ x r'_{a} +(1-x) r''_{J} } ~,
\label{amu1L}
 \end{align}
where $r'_a\equiv(M_{E_a}/m_\mu)^2$ and $r''_J\equiv(M_{S_J}/m_\mu)^2$.
%%%

%------------------------------------------------------------------------
\begin{figure}[t]
\centering
\includegraphics[width=5.5cm]{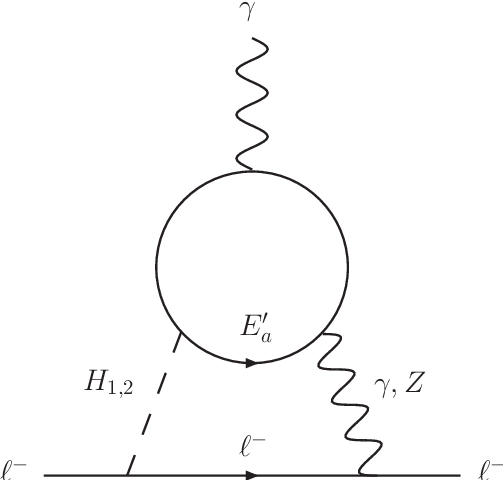}
\caption{A Barr-Zee diagram.}
\label{fig:BZ}
\end{figure}
%------------------------------------------------------------------------

As a subleading contribution, we have the Barr-Zee diagrams~\cite{Barr:1990vd,Cheung:2009fc} at the two-loop level, as depicted in Fig.~\ref{fig:BZ}. 
The relevant interactions are 
\begin{align}
\mathcal{L}_{H_i\bar{E}_aE_a}
&= -\sum_{i,a}H_i\bar{E}_a
\Big(
	g_{H_i\bar{E}_aE_a}^S+i\gamma_5g_{H_i\bar{E}_aE_a}^P
\Big)E_i,
\end{align}
where
\begin{align}
g_{H_1\bar{E}_aE_a}^S &= s_\alpha |y_{E_a}| c_{\phi_a} ~,\quad
g_{H_1\bar{E}_aE_a}^P = s_\alpha |y_{E_a}| s_{\phi_a} ~,
\\
g_{H_2\bar{E}_aE_a}^S &= c_\alpha |y_{E_a}| c_{\phi_a} ~,\quad
g_{H_2\bar{E}_aE_a}^P = c_\alpha |y_{E_a}| s_{\phi_a} ~,
\end{align}
where $s(c)_\alpha$ and $c(s)_{\phi_a}$ are the shorthand notations of $\sin(\cos)\alpha$, and $\cos(\sin)\phi_a$, respectively.

It is known that the $HZ$-type Barr-Zee diagram is accidentally suppressed by the $Z\bar{\mu}\mu$ coupling that is proportional to $(1/4-\sin^2\theta_W)\simeq 0.02$ with $\theta_W$ being the weak mixing angle. 
On the other hand, the $H\gamma$-type Barr-Zee contribution takes the form
 \begin{align}
& \Delta a_\mu^{(2)} \approx 
- \sum_{a=1}^{3}
\frac{\alpha_{\rm em}}{4\pi^3}\frac{m_\mu^2}{M_{E_a} v} |y_{E_a}|s_\alpha c_\alpha c_{\phi_a} 
\left[f(\tau_{a1})-f(\tau_{a2})\right],
\label{eq:subg2}
\end{align}
where
\begin{align}
f(\tau_{ai}) = 
\frac{\tau_{ai}}{2}\int_0^1dx~ \frac{1-2x(1-x)}{x(1-x)-\tau_{ai}}\ln\left(\frac{x(1-x)}{\tau_{ai}}\right),
\end{align}
with $\tau_{ai} \equiv (M_{E_a}/m_{H_i})^2$.
It should be noted that the negative coefficient of $f(\tau_{a2})$ is a consequence of the orthogonality of the rotation matrix $O(\alpha)$ defined in Eq.~(\ref{Omix}).  This implies that the Barr-Zee contributions would be highly suppressed provided $m_{H_1}\simeq m_{H_2}$.

{Even in the case of $m_{H_1}\neq m_{H_2}$,
$f(\tau_{ai})\simeq 13/18+(\ln \tau_{ai})/3$ in the limit $\tau_{ai}\gg 1$,} and Eq.~(\ref{eq:subg2}) is reduced to
{\begin{align}
\Delta a_\mu^{(2)} &\approx 
- \sum_{a=1}^{3}
\frac{\alpha_{\rm em}}{12\pi^3}\frac{m_\mu^2}{M_{E_a} v} |y_{E_a}|s_\alpha c_\alpha c_{\phi_a} 
\ln\left(\frac{m_{H_2}^2}{m_{H_1}^2}\right)
%%%
% ~\nn\\&\sim -10^{-10}
% \sum_{a=1}^{3}\frac{{\rm GeV}}{M_{E_a}} %|y_{E_a}| %c_{\phi_a} 
%\ln\left(\frac{m_{H_2}^2}{m_{H_1}^2}\right)
\nn\\&
\lesssim
-\text{sgn}(c_{\phi_a})(2.8\times 10^{-12})
\times
% \sum_{a=1}^{3}
\ln\left(\frac{m_{H_2}^2}{m_{H_1}^2}\right)
,\end{align}
where we have fixed $s_\alpha=0.1$, $|c_{\phi_a}| =1$, $|y_{E_a}|=1$ and $M_E=100$~GeV in the last line.
For $\text{sgn}(c_{\phi_a})=+1$, one must have $m_{H_2}< m_{H_1}(=125$ GeV) in order to generate 
the positive contribution, and it is the other way around for $\text{sgn}(c_{\phi_a})=-1$.
However, $\Delta a_\mu^{(2)}$ cannot reach the ${\cal O}(10^{-9})$ level in either case since the contribution
is logarithmic.  Moreover, the mixing angle $\alpha$ would vanish if the two masses are too far from each other.
Thus, the Barr-Zee contributions by themselves cannot be sufficiently sizeable to explain the muon $(g-2)$ anomaly, as will be shown in Sec.~\ref{sec:NA}.}

Since the couplings $F'_{ia}$ are generally complex, they can induce 
electric dipole moments (EDM's) for electron ($d_e$), neutron ($d_n$), and so on.
The current experimental upper bounds on $d_e$ and $d_n$ are respectively given by~\cite{Baron:2013eja}
\begin{align}
|d_e|<8.7\times10^{-29}~e~\text{cm}
~~\mbox{and}~~
|d_n| < 2.9\times 10^{-26}~e~\text{cm} ~.
\end{align}
In this model, the electron EDM imposes the strongest constraint on the CPV phases, so that we will focus 
on it.  We note in passing that the one-loop diagram is proportional to $|F'_{\ell a}|^2$ and hence does not induce the EDM's.  The nonzero contributions to $d_e$ are induced by the same Barr-Zee diagram
as in Fig.~\ref{fig:BZ}~\cite{Barr:1990vd}, and thus
\begin{align}
d_f=d_f^{H\gamma}+d_f^{HZ}.
\end{align}  
As in the muon $g-2$ case, the $HZ$-type Barr-Zee diagram is subdominant due to the accidentally suppressed $Z\bar{e}e$ coupling, and the $H\gamma$-type Barr-Zee contribution is cast into the form
\begin{align}
\frac{d_e^{H\gamma}}{|e|} =
\sum_{a=1}^{3}
\frac{\alpha_{\rm em}}{8\pi^3}\frac{m_e^2}{M_{E_a} v} |y_{E_a}|s_\alpha c_\alpha s_{\phi_a} 
\left[g(\tau_{a1})-g(\tau_{a2})\right] ~,
\label{de_Hgam}
\end{align}
where
\begin{align}
g(\tau_{ai}) = 
\frac{\tau_{ai}}{2}\int_0^1dx~ \frac{1}{x(1-x)-\tau_{ai}}\ln\left(\frac{x(1-x)}{\tau_{ai}}\right) ~.
\end{align}
As mentioned above, the two contributions of $H_{1,2}$ are destructive owing to the 
property of the orthogonal rotation matrix.
Since $g(\tau_{ai})\simeq 1+(\ln\tau_{ai})/2$ in the limit of $\tau_{ai}\gg1$, one gets
\begin{align}
\frac{d_e^{H\gamma}}{|e|} =
\sum_{a=1}^{3}
\frac{\alpha_{\rm em}}{16\pi^3}\frac{m_e^2}{M_{E_a} v} |y_{E_a}|s_\alpha c_\alpha s_{\phi_a} 
\ln\left(\frac{m_{H_2}^2}{m_{H_1}^2}\right). \label{eq:edm}
\end{align}

%%%%%%%%%%%%%%%%%%%%%%%%%%%%%%%%%%%%%%%%%%%%%%%%%%
\subsection{Signal Strengths of $H_1 \to \gamma\gamma$ Channel}
%%%%%%%%%%%%%%%%%%%%%%%%%%%%%%%%%%%%%%%%%%%%%%%%%%

%------------------------------------------------------------------------
\begin{figure}[t]
\centering
\includegraphics[width=8cm]{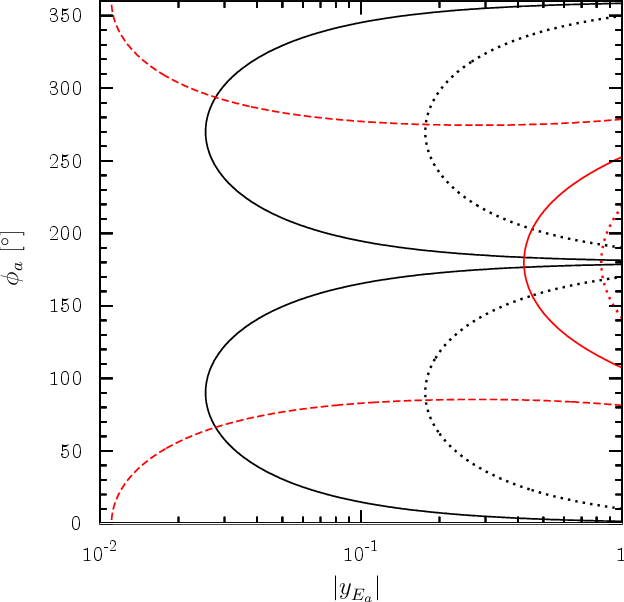}
\caption{Contours of $|d_e|=8.7\times 10^{-29}~e~{\rm cm}$ 
in the case of $m_{H_2}=500$ GeV (black solid curves) and 150 GeV (black dotted curves).
Regions to the right of each set of black curves are excluded by the electron EDM bound.
Also plotted are the Higgs diphoton signal strength $\mu_{\gamma\gamma}=0.9$ (red dashed curves), $1.0$ (red solid curve) and $1.1$ (red dotted curve), respectively.
Here, we take $c_\alpha=0.95$, $M_{E_1}=M_{E_2}=M_{E_3}=400$ GeV,
$|y_{E_1}|=|y_{E_2}|=|y_{E_3}|$ and $\phi_{1}=\phi_{2}=\phi_{3}$.}
\label{fig:eEDM}
\end{figure}
%------------------------------------------------------------------------

Due to the mixing between the two Higgs bosons, the couplings of $H_1$ with other SM particles are universally suppressed by the common factor $\cos\alpha$. 
\footnote{Although $H_1\to \tau\tau,~b\bar{b}$, etc can be modified by the doublet-singlet Higgs mixing, the current LHC data on them are not stringent enough to exclude the parameter space that we will explore below.}
However, the loop-induced $\gamma\gamma$ and $\gamma Z$ channels receive additional contributions from the exotic charged fermions, as seen from the Barr-Zee diagram in Fig.~\ref{fig:BZ}.  
Since the $\gamma Z$ mode has not been measured yet, we focus only on the $\gamma\gamma$
mode in what follows.  Nevertheless, the relative sizes of the deviations from the SM values in both modes are expected to be the same.

The signal strength of $H_1\to \gamma\gamma$ is approximately given by 
\begin{align}
\mu_{\gamma\gamma} 
& = 
\left[
	\bigg|c_\alpha+\frac{\mathcal{A}_E^S}{\mathcal{A}_{\rm SM}}\bigg|^2
	+\bigg|\frac{\mathcal{A}_E^P}{\mathcal{A}_{\rm SM}}\bigg|^2
\right]
\frac{c_\alpha^2\Gamma_{\rm SM}^{\rm tot}}{\Gamma^{\rm tot}_{H_1}} ~,
\label{mu2gam}
\end{align}
where
\begin{align*}
\Gamma^{\rm tot}_{H_1} 
&= c_\alpha^2\Gamma^{\rm tot}_{\rm SM}|_{\rm w/o\;\Gamma(H_1\to\gamma\gamma(Z))}
+\Gamma(H_1\to \gamma\gamma(Z))
+ \Gamma(H_1\to E_i^+E_j^-) \nn\\
&\quad+\Gamma(H_1\to S_RS_R)+\Gamma(H_1\to S_IS_I)+\Gamma(H_1\to \psi_i\psi_j) ~,
\end{align*}
$\mathcal{A}_{\rm SM}=-6.49$~\cite{Djouadi:2005gi}, 
$\Gamma_{\rm SM}^{\rm tot}\simeq4.1$~MeV~\cite{Heinemeyer:2013tqa}, and
$\mathcal{A}_E^{S,P}$ are respectively given by 
\begin{align}
\mathcal{A}_E^S 
&= \frac{2vs_\alpha|y_{E_a}|c_{\phi_a}}{M_{E_a}}\tau_a
\big\{1+(1-\tau_a)f_H(\tau_a)\big\} ~,~
\mathcal{A}_E^P
= \frac{2vs_\alpha|y_{E_a}|s_{\phi_a}}{M_{E_a}}\tau_a f_H(\tau_a) ~,
\end{align}
with $\tau_a \equiv 4M_{E_a}^2/m_{H_1}^2$ and the loop function $f_H(\tau)$ given 
in Ref.~\cite{Gunion:1989we}.
Assuming the dominance of SM contributions, Eq.~(\ref{mu2gam}) shows that the pseudoscalar couplings have minor effects on $\mu_{\gamma\gamma}$.

In the small $\alpha$ and large $M_{E}$ limit, one finds
\begin{align}
\mu_{\gamma\gamma}\simeq c_\alpha^2
\left[1+\frac{8v|y_{E_a}|c_{\phi_a}t_\alpha}{3M_{E_a}\mathcal{A}_{\rm SM}}\right] ~.
\label{mu2gam_app}
\end{align}
The deviation is mostly controlled by $c_\alpha^2$ rather than $\mathcal{A}_E^S$. 
Hence $\mu_{\gamma\gamma}$ is generally reduced in the model.

Since both $|d_e|$ and $\mu_{\gamma\gamma}$ are affected by the $E'_a$ loops, 
we briefly common on their correlations in the parameter space. 
Fig.~\ref{fig:eEDM} shows $|d_e|$ and $\mu_{\gamma\gamma}$
in the plane of $(|y_{E_a}|,\phi_a)$.  As a typical example, 
we set $c_\alpha=0.95$ and $M_{E_1}=M_{E_2}=M_{E_3}=400$ GeV, and assume
all the $|y_a|$ and $\phi_a$ are universal, respectively.
Contours of $|d_e|=8.7\times 10^{-29} ~e$ cm are plotted for $m_{H_2}=500$ GeV (black solid curves) 
and 150 GeV (black dotted curves).  Regions to the right of each set of black curves are excluded by the electron EDM limit at 90\% CL. 
The smaller $m_{H_2}$ case is less sensitive to the electron EDM because of the cancellation mechanism 
at work, as can be seen from Eq.~(\ref{de_Hgam}), thereby allowing more parameter space.

As for the Higgs diphoton signal strength, we display $\mu_{\gamma\gamma}=0.9$ (red dashed curve), $1.0$ (red solid curve) and $1.1$ (red dotted curve), respectively.  As mentioned above, $\mu_{\gamma\gamma}$ is less than unity in most parameter space, which is due mainly to the factor of $c_\alpha^2$.  However, the loop effects of $E'_a$ can be constructive to the SM contribution for $c_{\phi_a}<0$, and render $\mu_{\gamma\gamma}\ge1$ if $|y_{E_a}|\gtrsim 0.5$.

%%%%%%%%%%%%%%%%%%%%%%%%%%%%%%%%%%%%%%%%%%%%%%%%%%
\subsection{Flavor-Changing Leptonic $Z$ Boson Decays}\label{subsec:Zll}
%%%%%%%%%%%%%%%%%%%%%%%%%%%%%%%%%%%%%%%%%%%%%%%%%%

%------------------------------------------------------------------------
\begin{figure}[t]
\centering
\includegraphics[width=5cm]{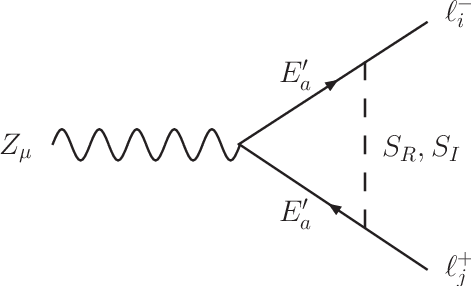}
\hspace{0.3cm}
\includegraphics[width=5cm]{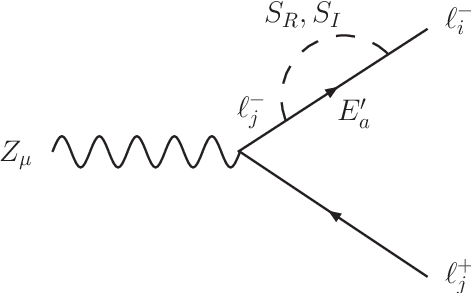}
\hspace{0.3cm}
\includegraphics[width=5cm]{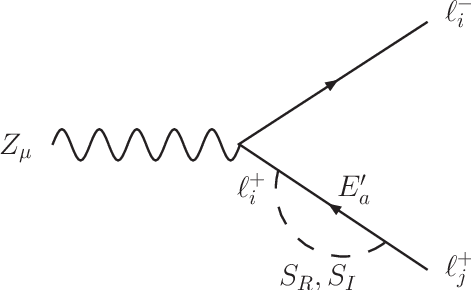}
\caption{One-loop contribution to the $Z\to \ell_i^- \ell_j^+$ decay. }
\label{fig:Zll}
\end{figure}
%------------------------------------------------------------------------

Here we consider the decay of the $Z$ boson to two charged leptons of different flavors at the one-loop level, as shown in Fig.~\ref{fig:Zll}.  The amplitudes of such decay modes involve the Yukawa couplings $F'_{ia}$, some of which can be of ${\cal O}(1)$ in order to achieve a sizeable contribution to the muon $g-2$. 
 %%%
After summing up the three diagrams, the UV divergences cancel out and the finite part is cast into the form
\begin{align}
\text{BR}(Z\to\ell^-_i\ell^+_j)
&=
\frac{G_F}{3\sqrt2 \pi} \frac{m_Z^3}{(16\pi^2)^2 \Gamma_{Z}^{\rm tot}} 
\left(s_W^2 -\frac12\right)^2
\nn\\
& \qquad\qquad
\times
\left| \sum_{a=1}^{3} \sum_{J = R,I} F'_{ia} {F'_{ja}}^* 
\left[ F_2(E_a,S_J)+F_3(E_a,S_J) \right] \right|^2 ~,
\label{eq:Zll}
\end{align}
where
\begin{align*}
F_2(a,b) &=\int_0^1dx(1-x)\ln\left[ (x m_a^2 + (1-x) m_b^2 \right] ~,
\\
F_3(a,b) &=\int_0^1dx\int_0^{1-x}dy\frac{(2xy m_Z^2+(m_a^2-m_b^2)(1-x-y)-\Delta\ln\Delta}{\Delta} ~,
\end{align*} 
with $\Delta\equiv -xy m_Z^2+(x+y)(m_a^2-m_b^2)+m_b^2$ and the total $Z$ decay width $\Gamma_{Z}^{\rm tot} = 2.4952 \pm 0.0023$~GeV~\cite{pdg}.
From Eqs.~(\ref{eq:damu1}) and (\ref{eq:Zll}), one can see that the FCNC couplings 
$F'_{ia} {F'_{ja}}^*$ identically appear in $\ell_i\to\ell_j\gamma$ and $Z\to \ell_i \ell_j $, 
and hence they can be correlated with each other.
However, one crucial difference is their decoupling properties. 
{The former modes would be suppressed as the particles in the loops become heavy while the latter can grow logarithmically.  This difference may stem from the different structures in the form factors: the former of the dipole type and the latter of the vector one.  A similar nondecoupling behavior of the LFV $Z$ decays can be found in Ref.~\cite{Illana:2000ic}, where $\text{BR}(Z\to\mu\tau)$ can grow with the quartic power of an internal particle mass.
}

The current lepton flavor-changing $Z$ boson decay branching ratios are found to be~\cite{pdg}:
\begin{align}
\begin{split}
  {\rm BR}(Z\to e^\pm\mu^\mp) &< 1.7\times10^{-6} ~,\\
  {\rm BR}(Z\to e^\pm\tau^\mp) &< 9.8\times10^{-6} ~,\\
  {\rm BR}(Z\to \mu^\pm\tau^\mp) &< 1.2\times10^{-5} ~,\label{eq:zmt-exp}
\end{split}
\end{align}
where the upper bounds are quoted at 95 \% CL.
We have scanned the parameter space and found that all these constraints are less stringent than those from the LFV processes,
{as well as the flavor-conserving processes ${\rm BR}(Z\to \ell^\pm\ell^\mp)$ ($\ell=e,\mu,\tau$). }

%%%%%%%%%%%%%%%%%%%%%%%%%%%%%%%%%%%%%%%%%%%%%%%%%%
\subsection{Dark Matter Candidates}
%%%%%%%%%%%%%%%%%%%%%%%%%%%%%%%%%%%%%%%%%%%%%%%%%%

In our model, we have both bosonic $S_{R(I)}$ and fermionic $\psi_1$ DM candidates, which will be generically denoted by $X$. 
To analyze each of the two scenarios, we simply assume that any quartic couplings and trilinear couplings involving the DM candidate after the EW symmetry breaking are negligibly small except for the quartic couplings that are required to be sufficiently larger in order to retain the vacuum stability.
In the case of the bosonic DM candidate, it is easy to evade the constraints of direct detection searches.
Moreover, we focus on the DM mass regime of $1~{\rm GeV}\lesssim M_X\lesssim 100$~GeV.
As a consequence, the $X \to H_1 H_1$ decay is kinematically forbidden.

In our numerical analysis, we will take a somewhat relaxed range of $0.11\lesssim \Omega h^2\lesssim 0.13$ in
comparison with the the one reported by Planck Collaboration, $\Omega h^2\approx 0.12$~\cite{Ade:2013zuv}. 

%------------------------------------------------------------------------
\begin{figure}[t]
\centering
\includegraphics[width=6cm]{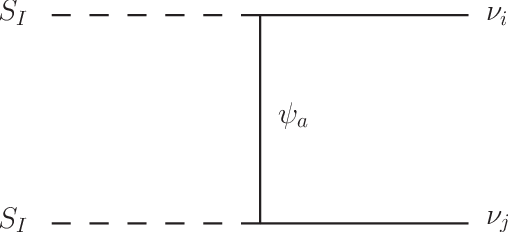}
\caption{Dominant annihilation cross section in the bosonic DM scenario. }
\label{fig:BDM}
\end{figure}
%------------------------------------------------------------------------

{\it Bosonic DM}:

We first consider the bosonic DM candidate $S_I$.  The case of having $S_R$ as the lighter scalar boson and DM candidate is phenomenologically the same.
{The most stringent upper bound on the spin-independent DM-nucleon scattering cross section reported by the LUX experiment~\cite{Akerib:2016vxi} is $\sigma_N\lesssim 2.2\times 10^{-46}$~cm$^2$ at around $M_X=50$~GeV.
We will adopt this upper bound for the entire range of DM mass considered in this work 
for simplicity. 
%Even in such a case, the direct detection constraint is easily evaded 
%in the both bosonic and fermionic DM cases, where the latter will be discussed below.
%Notice here that the fermionic DM does not have the {\color{red}sizable spin dependent} contribution, since it is Majorana type.
%Thus we consider the bound on bosonic DM candidate only.

The cross section of $S_I$ scattering with a nucleon is given by
\begin{align}
\sigma_{\rm SI}(S_I N\to S_I N)&\approx
\left| \frac29+\frac79\sum_{q=u,d,s}f_q\right|^2
\frac{\lambda_{\Phi S}^2 m_N^4}{4\pi (m_{S_I} + m_N)^2 m_{H_1}^4}\nn\\
&\approx(3.29\times10^{-29}~{\rm cm^2})\times
\frac{\lambda_{\Phi S}^2 m_N^4}{4\pi (m_{S_I} + m_N)^2 m_{H_1}^4}
%\left[\lambda_{\Phi S}v 
% \left(\frac{c_\alpha^2}{m_{H_1}^2} + \frac{s_\alpha^2}{m_{H_2}^2}\right)
%+(\lambda_{\Phi \varphi}v +\mu_{S_1}) \left(\frac{1}{m_{H_1}^2} - \frac{1}{m_{H_2}^2}\right) \right]
, 
\end{align}
where $m_N\approx 0.939$ GeV is the neutron mass, and we assume that $m_{H_1}=m_{H_2}$ for simplicity.
%The value of $3.30\times10^{-29}$ cm$^2$ is computed by the result of lattice simulation, and find to be
%\begin{align}
%\left| \frac29+\frac79\sum_{q=u,d,s}f_q\right|^2\frac{1}{{\rm GeV^{2}}}\approx 0.287^2\times \frac{\rm cm^2}{2.5\times10^{27}},
%\end{align}
In the second line, $f_u=0.0110$, $f_d=0.0273$ and $f_s=0.0447$ are used.
%$s_\alpha=0.1$, 
For $m_{S_I}=50$~GeV, one finds an upper bound on $\lambda_{\Phi S}$:
\begin{align}
\lambda_{\Phi S}\lesssim 0.0083.
\end{align}
We can always choose $\lambda_{\Phi S}$ that satisfies this bound without 
affecting other phenomenological discussions. 
}

{As shown in Fig.~\ref{fig:BDM}}, the dominant annihilation cross section that affects the DM relic density derives from the $f_{ij}$ couplings between the {neutrinos} and exotic fermions.  
Written in the mass eigenbasis, the scattering cross section is given by
\footnote{We have confirmed that the $\mathcal{O}(v_{\rm rel}^2)$ term in the annihilation cross section is so small that it does not affect our conclusions below.}
\begin{align}
\sigma v_{\rm rel}\approx \sum_{a=1}^9 \sum_{i,j=1}^3\frac{|F_{ia} F^T_{aj}|^2 M_a^2}{4\pi (M_a^2+M_X^2)^2}+{\cal O}(v_{\rm rel}^2) ~.
\label{BDMrelic}
\end{align}
This shows that the DM annihilation to a pair of neutrinos is dominantly $S$-wave, a consequence of the $t$- and $u$-channel mediators being Majorana particles.
%%%
The relic density $\Omega h^2$ is then given by~\cite{Srednicki:1988ce}
\begin{align}
\Omega h^2\approx \frac{1.07\times 10^{9} x_f }{\sqrt{g_*(x_f)} M_P  a_{\rm eff} }
~~\mbox{with}~~
a_{\rm eff} =  \sum_{a=1}^9 \sum_{i,j=1}^3\frac{|F_{ia} F^T_{aj}|^2 M_a^2}{4\pi (M_a^2+M_X^2)^2} ~,
\label{eq:relic}
\end{align}
where the Planck mass $M_P\approx 1.22\times10^{19}$~GeV, $g_*(x_f\approx25) \approx 100$ is the total number of effective relativistic degrees of freedom at the time of freeze-out, and $x_f\approx25$ is defined by $M_X/T_f$ at the freeze-out temperature $T_f$.

{The only currently available possibility to detect the bosonic DM indirectly is the IceCube experiment~\cite{Aartsen:2014gkd},
since the bosonic DM's annihilate into neutrinos.
However it requires that the DM have a large cross section and a mass at the PeV scale, which is far beyond the DM mass range of interest to us.
}

{\it Fermionic DM}:
In the case of a fermionic DM, the lightest one of the nine $\psi_a$ bosons may not be a DM candidate.
This is because a neutral fermion originated from the gauge doublet $N'$ cannot be a DM candidate, as it has been ruled out by the direct detection searches via the $Z$ boson portal.  Hence, only the lightest one of the gauge singlet fermion $N$ can be a DM candidate.
Here we assume $m_{LR}\approx0$ for simplicity and, as a consequence, do not need to worry about the $Z$ portal due to the mixing between $N$ and $N'$.  Nevertheless, we still have to take into account the Higgs portal as another channel for the direct detection constraint.
The spin independent cross section between the lightest gauge singlet $N$ and the nucleon mediated by the two Higgs bosons is given by
\begin{align}
\sigma_{N}\approx (3.29\times 10^{-29}~{\rm cm^2})\times \frac{\mu_X^2 \text{Re}[(y_{N})_{11}]^2 m_N^2 s_\alpha^2 c_\alpha^2}{\pi v^2} \left|\frac{1}{m_{H_1}^2}-\frac{1}{m_{H_2}^2}\right|^2 ~,
\end{align}
{where 
$\mu_X \equiv M_X m_N / (M_X+m_N)$ is the reduced mass.
For $\text{Re}[(y_{N})_{11}]\simeq 0$ or $m_{H_1}\simeq m_{H_2}$, 
the spin-independent DM cross section is highly suppressed.
As in the cases of the muon $g-2$ and the electron EDM at the two-loop level
(see Eqs.~(\ref{eq:subg2}) and (\ref{de_Hgam})), 
the destructive interference between the two contributions 
is a direct result of the orthogonality of the rotation matrix $O(\alpha)$.
The importance of such a cancellation in the spin-independent DM cross section is emphasized 
in Refs.~\cite{Kim:2008pp,Baek:2011aa} (see also Refs.~\cite{Baek:2012uj,Baek:2012se}).}

%------------------------------------------------------------------------
\begin{figure}[t]
\centering
\includegraphics[width=6cm]{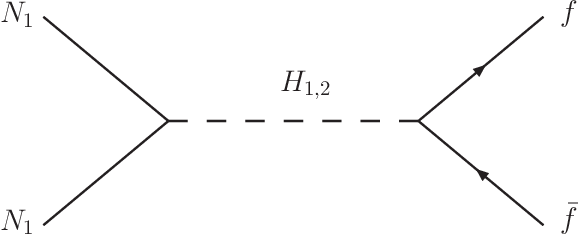}
\hspace{0.5cm}
\includegraphics[width=6cm]{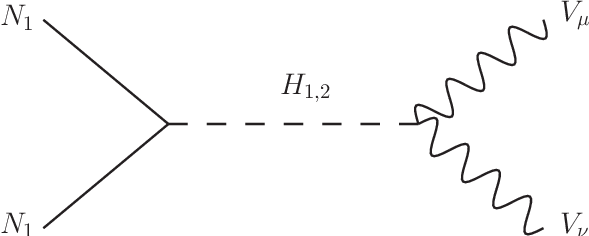}
\caption{Dominant annihilation cross section in the fermionic DM scenario. }
\label{fig:FDM}
\end{figure}
%------------------------------------------------------------------------

The cross section of the DM relic density arises from the interactions involving the $y_N$ couplings
{as shown in Fig.~\ref{fig:FDM}}, and its form is given by
\begin{align}
(\sigma v_{\rm rel})
& \approx \sum_f\frac{N_C^fm_f^2s_\alpha^2 c_\alpha^2 s\beta_f^3}{8\pi v^2}
\Big[
	\text{Re}[(y_N)_{11}]^2\beta_{X}^2
	+\text{Im}[(y_N)_{11}]^2
\Big]|G|^2 \nn \\
&\quad +  \sum_{V=Z,W}\frac{S_Vm_V^4s_\alpha^2 c_\alpha^2\beta_V }{4\pi v^2}
\Big[
	\text{Re}[(y_N)_{11}]^2\beta_{X}^2
	+\text{Im}[(y_N)_{11}]^2
\Big]\left[3+\frac{s^2}{4m_V^4}\beta_V^2\right]|G|^2, 
\label{eq:sigmavFermion1}
\end{align}
where $\beta_F = \sqrt{1-4M_F^2/s}~~(F= f,\ X,\ V)$ and 
\begin{align}
G &= \frac{1}{s-m_{H_1}^2+im_{H_1}\Gamma_{H_1}}
- \frac{1}{s-m_{H_2}^2+im_{H_2}\Gamma_{H_2}},
\label{eq:sigmavFermion}
\end{align}
with $s$ being the Mandelstam variable, the summation of $f$ running over all the SM fermions, $N^f_C=1$ for leptons, $N^f_C=3$ for quarks, and $S_V = 1/2$ (1) for $V=Z$ ($W$).  In Eq.~\eqref{eq:sigmavFermion}, the first term includes the SM fermion pairs, and the second one the SM weak gauge boson pairs.

{From Eqs.~(\ref{eq:sigmavFermion1}) and (\ref{eq:sigmavFermion}), 
one can see that $\sigma v_{\rm rel}\propto M_X^2/m_{H_1}^4$ for $M_X\ll m_{H_1}\ll m_{H_2}$,
and $\sigma v_{\rm rel}\propto 1/M_{X}^2$ for $m_{H_1}\ll M_X \ll m_{H_2}$.
On the other hand, $H_2$ comes into play if $m_{H_2}\simeq m_{H_1}$ or $m_{H_2}\simeq M_X$.
For instance, there would be a partial cancellation between the $H_1$ and $H_2$ contributions
for $m_{H_1}\simeq m_{H_2}$.
Furthermore, $\sigma v_{\rm rel}$ would be resonantly enhanced
if $M_X\simeq m_{H_1}/2$ or $m_{H_2}/2$.}

The total decay width of $H_1$ is modified when the $H_1 \to XX$ channel is open, and that of $H_2$ is dominated by $\Gamma_{H_2\to 2X}$.  That is,
\begin{align}
&
\Gamma_{H_1} \approx c_\alpha^2 \Gamma_{\rm SM}^{\rm tot} + \Gamma_{H_1\to 2X}
~~\mbox{and}~~
\Gamma_{H_2}\approx \Gamma_{H_2\to 2X} ~,
\nn \\
&
\mbox{with}~
\Gamma_{H_i\to 2X} =
\frac{m_{H_i} O_{2i}^2 }{16\pi} \sqrt{1-\frac{4 M_X^2}{m_{H_i}^2}} 
\left[ {\rm Re}[ (y_{N})_{11} ]^2\left(1-\frac{4 M_X^2}{m_{H_i}^2}\right) 
+ {\rm Im}[ (y_{N})_{11} ]^2 \right]
\end{align}
for $i = 1,2$.
We expect $\Gamma_{H_i} \ll m_{H_i}$ ($i=1,2$), 
%%%%%%%
{and} the relic density of DM is given by
\begin{align}
&\Omega h^2
\approx 
\frac{1.07\times10^9}{\sqrt{g_*(x_f)}M_{P} J(x_f)},
\label{eq:relic-deff}
\end{align}
where again $g^*(x_f\approx25)\approx100$
and $J(x_f)$ is given by~\cite{Ko:2017yrd, Edsjo:1997bg}
\begin{align}
&
J(x_f)=\int_{x_f}^\infty dx
\left[ \frac{\int_{4M_X^2}^\infty ds\sqrt{s-4 M_X^2} (\sigma v_{\rm rel}) 
K_1\left(\frac{\sqrt{s}}{M_X} x\right)}
{16  M_X^5 x K_2(x)^2}\right] ~,
\label{eq:relic-deff}
\end{align}
where $K_{1,2}$ are the modified Bessel functions of the second kind 
of order 1 and 2, respectively.
%%%%%%%
We find that the solution to obtain a sizeable muon $g-2$ correction is at around half the mass of the mediating particle.
Therefore, we fix $M_X \approx m_{H_1}/2 \approx 62.5$~GeV and close the $H_1 \to X X$ channel.~\footnote{Although we have another solution $M_X\approx m_{H_2}/2$, $M_X\approx m_{H_1}/2$ is more promising for direct detection.  Thus, we focus on this solution.  Note also that the direct detection bound is more stringent than the invisible decay of the SM Higgs boson at this scale.}
Notice that here we have to apply the exact formula Eq.~(\ref{eq:relic-deff}), which is unlikely to the case of bosonic DM, to get the correct relic density at around the pole, integrating $s$ from $4M_X^2$ to infinity.
%%%
Furthermore, we fix $m_{H_2} = 150$~GeV and $s_\alpha\approx0.1$ for numerical analyses.  We then find that the upper bound on $|(y_{N})_{11}|^2$ is
$0.81$ from the direct detection searches.  When using $|(y_{N})_{11}|^2 = 0.81$, we further obtain $\Gamma_{H_2} \approx 1.32$~GeV, much less than $m_{H_2}$, while $\Gamma_{H_1}$ is virtually the same as the SM value.  Therefore, the resonance condition $M_X \approx m_{H_1}/2\approx 62.5$~GeV provides a sufficient enhancement for the DM annihilation cross section to render the desired DM relic density.
In the above numerical estimation, ${\rm Re}(y_{N})_{11}$ plays a much less significant role in the determination of Higgs boson widths and DM annihilation rate.  We therefore take it to be 0 for simplicity.

{It is worth considering the indirect DM detection via extra photon emissions, as measured and reported by the Fermi-LAT experiment.
For example, the monochromatic anomaly of a DM of mass 43~GeV and an annihilation cross section of ${\cal O}(10^{-11})$ GeV$^{-2}$~\cite{Liang:2016pvm}
might be realized by judiciously tuning $m_{H_2}$ in the current model.}

%%%%%%%%%%%%%%%%%%%%%%%%%%%%%%%%%%%%%%%%%%%%%%%%%%%%%%%%%%%%%%%%%%%%%%%%%%%%%%%%%%%%%%%%%%%%%%%%%%%%%%%%%

%%%%%%%%%%%%%%%%%%%%%%%%%%%%%%%%%%%%%%%%%%%%%%%%%%
\section{Numerical analysis \label{sec:NA}}
%%%%%%%%%%%%%%%%%%%%%%%%%%%%%%%%%%%%%%%%%%%%%%%%%%

In this section, we present our results in the exploration of allowed parameter space that satisfies all the constraints discussed in the previous section.  We concentrate on the region in which we can simultaneously obtain a sizeable muon $g-2$ toward an explanation for the observed anomaly and have a bosonic or fermionic DM candidate.  In such an exercise, we fix the Higgs boson mixing angle to have $s_\alpha = 0.1$.  The value of electron EDM is predicted at around $10^{-30}\sim 10^{-28} ~e$ cm, close to the current experimental upper bound.

As alluded to before, we take $m_{LR}$, $M_N$, $V_C$ and $V_N$ to be diagonal for simplicity in our numerical analyses.  Our findings have little dependence on these assumptions.

{\it Bosonic DM Case}:
Before delving into a detailed discussion of the bosonic DM case, we remind the reader that the mass matrix of exotic neutral fermions can be assumed to have $M_{E},m_{LR} \ll M_{N}$, as only the smallest three masses are constrained by the active neutrino oscillation.  As confirmed numerically, the matrix $M_N$ can take any sufficiently large values without affecting our results.
Thus, we can take any (large) mass eigenvalues for the six heavy exotic neutral fermions, and realize baryogenesis via a high-scale leptogenesis as described below.

%%%%%%%%%%%%%%%%%%%
\begin{figure}[t]
\begin{center}
\includegraphics[width=120mm]{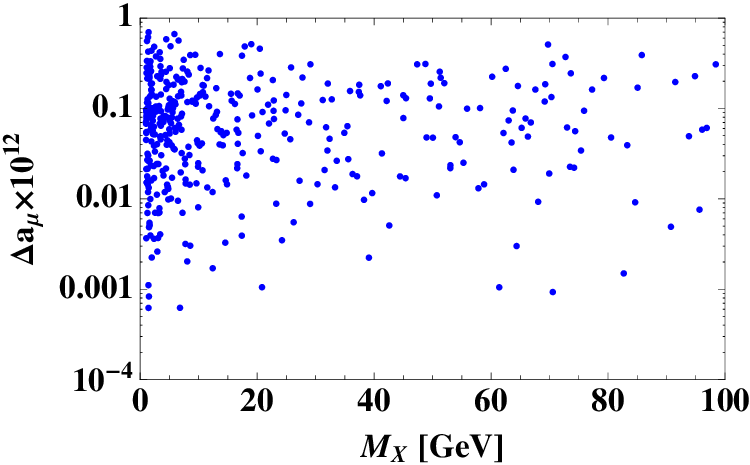} \qquad
\caption{Scatter plot of allowed ranges of $M_X$ and $\Delta a_\mu \times 10^{12}$ for the bosonic DM case.
{It is found that $\Delta a_\mu$ cannot reach 
$\mathcal{O}(10^{-9})$ in the bosonic DM case.
Such a small muon $g-2$ is due to the fact that the couplings $F'$ appearing in (\ref{amu1L})
are constrained by the upper bound of $F$, as determined by the observed DM relic density.  
}
}
  \label{fig:DM-damu}
\end{center}\end{figure}
%%%%%%%%%%%%%%%%%%%

Explicitly, we scan the following parameter ranges:
\begin{align}
&
 0\lesssim (\text{Re}(\delta_{23}) , \text{Re}(\delta_{13}) , \text{Re}(\delta_{12})) \lesssim \pi ~,~
 0.1\lesssim (\text{Im}(\delta_{23}) , \text{Im}(\delta_{13}) , \text{Im}(\delta_{12})) \lesssim 10  ~,~
 \nn\\
 &
 |\phi_{1,2,3}| \lesssim 2\pi ~,~
 |y_{E_{1,2,3}}| \lesssim 1 ~, ~\nn\\
& 
 M_X \lesssim 100 {\rm GeV}~,~ 
 m_{H_1}\lesssim m_{H_2} \lesssim 600\ {\rm GeV}~,~
 1.2 M_X \lesssim m_{S_R} \lesssim 300\ {\rm GeV}~,\nn\\
& 
%%%
 {100}\ {\rm GeV} \lesssim M_{E_1}  \lesssim 500\ {\rm GeV}~,~  
 M_{E_{1(2)}}  \lesssim M_{E_2}(M_{E_3}) \lesssim 2000 \ {\rm GeV}~,
\nn\\
& 
 1.2 M_{X} \lesssim m_{LR_1} \lesssim 500 \ {\rm GeV}~,~  
 m_{LR_{1(2)}} \lesssim m_{LR_2}(m_{LR_3}) \lesssim 2000 \ {\rm GeV}~.
\end{align}
{Here each range of $(\delta_{23} , \delta_{13} , \delta_{12})$ is the typical scale to fit the neutrino oscillation data.}
Moreover, we take the perturbativity limit as $4\pi$ for all the $f_{ij}$ couplings.
We have randomly prepared $10^6$ points in the above-mentioned parameter space, and found that ${\it 360}$ of them pass all the constraints, where we neglect the negative solutions of muon $g-2$.  Fig.~\ref{fig:DM-damu} is a scatter plot showing the DM mass and $\Delta a_\mu$ of these allowed parameter sets.  Although the allowed $M_X$ spans over the entire range of interest to this work, the muon $g-2$ is at most of ${\cal O}(10^{-12})$, far less than the required $\Delta a_\mu= {\cal O}(10^{-9})$.
%%%
{It should be noted that the couplings $F'$ appearing in the $g-2$ formula (\ref{amu1L}) are related to the couplings
$F$ entering the DM relic density (\ref{BDMrelic}) via $F'=FV_N$ 
under the currently adopted texture of $M$ defined in Eq.~(\ref{eq:neutral-mat}).
For most of the scanned parameter space, it turns out that the couplings $F$ are small,
and thus the points in Fig.~\ref{fig:DM-damu} are denser in the region where $M_X\lesssim 10$ GeV 
to be consistent with the observed DM relic density. 
Our scan analysis shows that $F'$ are not allowed to exceed ${\cal O}(0.01)$ 
in order to avoid the over-abundant DM relic density.  This in turn prevents $\Delta a_\mu$ from being sufficiently large.
Note that if the DM annihilation cross section is dominated by the $P$-wave rather than the $S$-wave,
the couplings $F$ could be larger.
}

{Here, we also comment on some experimental constraints from LFV processes.
The strongest one comes from $\text{BR}(\mu\to e\gamma)<4.2\times 10^{-13}$.
However, it could be evaded if the couplings $F'$ take specific forms. 
Focusing on the dependence of $F'$ in muon $g-2$, one finds
\[
\Delta a_\mu \propto |F_{21}'|^2 +  |F_{22}'|^2 +  |F_{23}'|^2 ~,
\]
while
\[
{\rm BR}(\mu\to e \gamma) \propto F'_{11} F_{21}^{'*} +  F'_{12} F_{22}^{'*} +F'_{13} F_{23}^{'*} ~.
\]
Thus, there should be some parameter space where $F_{21}'$, $F_{22}'$ and $F_{23}'$ are large
while $F_{11}'$, $F_{12}'$ and $F_{13}'$ are small enough to satisfy the constraint of $\mu\to e \gamma$.
However, one should note that the texture of $M$ as well as the relation $F'=FV_N$ 
do not always grant such a region, and the bosonic DM scenario presented here is indeed the case. 
We will see a working case in the fermionic DM scenario below.
}

Before moving on to the fermionic DM case, we comment on a possibility of leptogenesis. 
In the standard high-scale leptogenesis, CP violation arises from the vertex of $\bar L_{L}N_{R}\tilde{\Phi}$, and the decays of $N_R$ generate a lepton asymmetry which is eventually  
converted to the baryon asymmetry through a sphaleron process~\cite{Fukugita:1986hr}. 
In our model, however, such a term is forbidden by the $Z_2$ symmetry.
Nevertheless, owing to the similar term $g \bar L'_{L}N_{R}\tilde{\Phi}$, 
the lepton asymmetry may still arise by the decays of $N_R$. 
The CPV parameter in this case is  
\begin{align}
\epsilon_i &= \frac{\sum_j\big[\Gamma(N_i\to L'_j \phi)-\Gamma(N_i\to \bar{L}'_j \bar{\phi})\big]}
{\sum_j\big[\Gamma(N_i\to L'_j \phi)+\Gamma(N_i\to \bar{L}'_j \bar{\phi})\big]} \nn\\
&= \frac{1}{8\pi} \frac{1}{(g^\dagger g)_{ii}}
\sum_{k\neq i} {\rm Im}\big[(g^\dagger g)^2_{ii}\big]\Big[f(\xi_k)+g(\xi_k)\Big],
\end{align}
with $\xi_k=M_{N_k}^2/M_{N_i}^2$ and 
\begin{align}
f(\xi) = \sqrt{\xi}\left[1-(1+\xi)\ln\frac{1+\xi}{\xi}\right], \quad
g(\xi) = \frac{\sqrt{\xi}}{1-\xi}.
\end{align}
Here, the masses of $\phi$ and $L'$ are neglected.
It should be noted that unlike the ordinary case, the coupling $g$ is not restricted by the 
low-energy neutrino data, giving rise to a sufficient CP asymmetry.
{Since an estimate of the final baryon number density is highly model dependent, 
the detailed analysis will be given elsewhere.
In contrast to the bosonic DM scenario, the above leptogenesis would not work in the fermionic DM scenario,
where the right-handed neutrino is the DM candidate, as discussed below.}

{\it Fermionic DM Case}:
First of all we fix $m_{LR} = 0$ for simplicity.  The condition avoids the possibility of mixing between the gauge singlet and doublet.  We therefore do not need to worry about the more stringent constraint from spin-independent DM-nucleon scattering via the $Z$ boson portal.
In this case, the lightest entry of $M_N$ is automatically identified as the DM mass. 
%%%
Secondly, we fix ${\rm Re}(y_N)_{11} = 0$ and ${\rm Im}(y_N)_{11} = 0.9$ as given by the most conservative bound from the direct detection searches.  We then obtain $\Gamma_{H_1} \approx 0.0041$~GeV and $\Gamma_{H_2} \approx 1.32$~GeV.  We also take the resonance condition $M_X \approx 62.5$~GeV so as to get the correct relic density $\Omega h^2\approx 0.12$. 
%%%
{We further make an assumption of mass degeneracy: $ M_{E_1} \approx m_{S_R} \approx m_{S_I}$.  It plays a crucial role in obtaining a sizeable muon $g-2$ due to the loop function in Eq.~(\ref{eq:damu1}).~\footnote{One can readily check that the loop function becomes very small if there is a big mass difference among them.  Here we take these mass differences to be of order $10^{-5}\sim10^{-3}$ and $10^{-11}\sim 10^{-8}$~GeV, respectively.}

In addition to the above assumptions, we further take $M_E\equiv M_{E_i}$, $\phi\equiv \phi_i$, $y_E\equiv y_{E_i}$ $(i=1,2,3)$ for simplicity.
We scan the following parameter ranges:
\begin{align}
&
 0\lesssim (\text{Re}(\delta_{23}) , \text{Re}(\delta_{13}) , \text{Re}(\delta_{12})) \lesssim \pi ~,~
 0.1\lesssim (\text{Im}(\delta_{23}) , \text{Im}(\delta_{13}) , \text{Im}(\delta_{12})) \lesssim 10  ~,~
 \nn\\
& |\phi_{1,2,3}| \lesssim 2\pi ~,~0.01\lesssim  |y_{E}| \lesssim 1 ~, \nn\\
& ~m_{H_1} \lesssim {m_{H_2}} \lesssim 500  \ {\rm GeV}~,~ 
200 \ {\rm GeV} \lesssim {M_{E}} \lesssim 1000 \ {\rm GeV} ~,~\nn \\
&
%%% 
 1.2 M_{X}  \lesssim {M_{N_2}} \lesssim 1000 \ {\rm GeV}~,~  
 M_{N_{2}}  \lesssim {M_{N_3}} \lesssim 1500  \ {\rm GeV}~.
\end{align}
Moreover, we take the perturbativity limit as $4\pi$ for all the $f_{ij}$ couplings.

We have randomly prepared $1.5\times10^6$ points in the above-mentioned parameter space, and found that ${\it 630}$ of them pass all the constraints, including $1.5\times 10^{-9}\lesssim \Delta a_\mu\lesssim 4.0\times 10^{-9}$.

%%%%%%%%%%%%%%%%%%%
\begin{figure}[h!]
\begin{center}
\includegraphics[width=100mm]{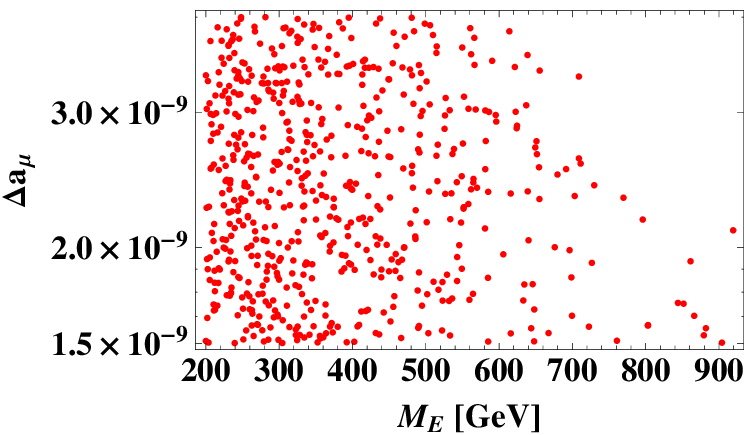} 
\\%[0.5cm]
\vspace{10pt}
\includegraphics[width=100mm]{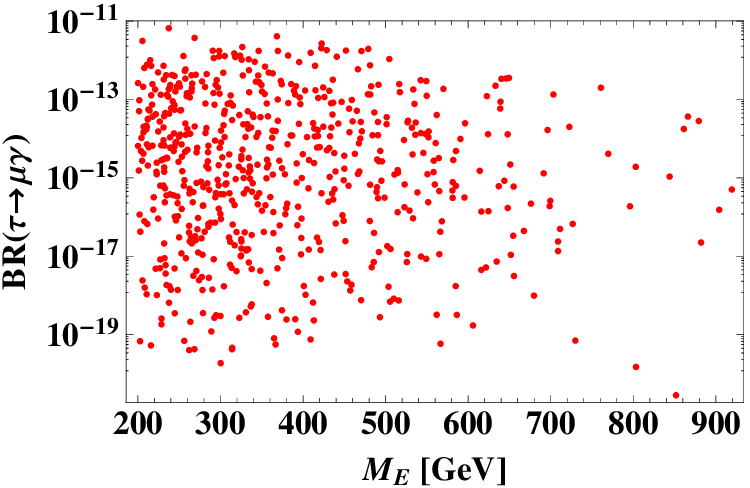}
\\
\vspace{10pt}
\includegraphics[width=100mm]{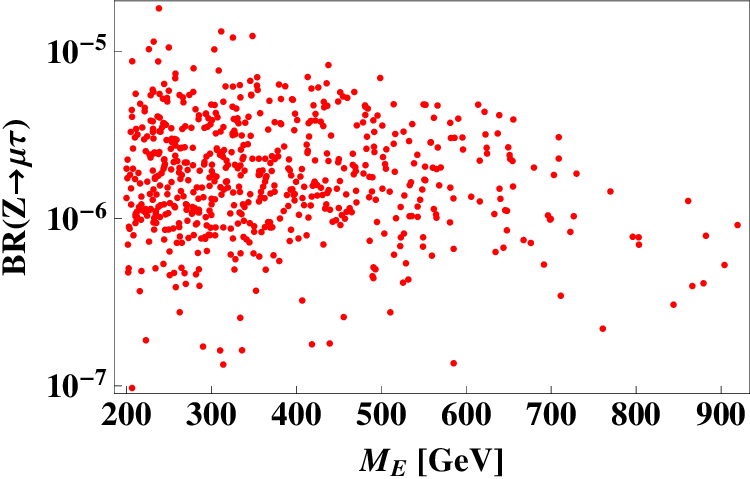}
\\
\caption{Scatter plots of allowed ranges for $\Delta a_\mu$ (top plot), $\text{BR}(\tau\to\mu\gamma)$ (middle plot) and $\text{BR}(Z\to\mu\tau)$ (lower plot) as a function of $M_{E}$, satisfying $1.5\times 10^{-9}\lesssim \Delta a_\mu\lesssim 4.0\times 10^{-9}$. These figures indicates an upper bound on $M_E$ of around 1~TeV, which comes from the constraint of neutrino oscillation data. }
  \label{fig:FDM-ME}
\end{center}\end{figure}
%%%%%%%%%%%%%%%%%%%
 Fig.~\ref{fig:FDM-ME} shows the scatter plots of allowed ranges for the muon $g-2$ (top), $\text{BR}(\tau\to\mu\gamma)$ (middle), $\text{BR}(Z\to\mu\tau)$ (lower) as a function of $M_{E}$.  
 The muon $(g-2)$ and $\text{BR}(\tau\to\mu\gamma)$ would be suppressed with increasing 
 $M_{E}$ as expected.
However, $\text{BR}(Z\to \mu\tau)$ can in principle grow as $M_{E}$ increases
owing to the nondecoupling property, as mentioned in Sec.~\ref{subsec:Zll}.
The suppression of $\text{BR}(Z\to \mu\tau)$ observed here 
actually comes from the suppression of the 
FCNC couplings that are controlled by the neutrino mass generation.
%%%%%%%%%%%%%%%%%%%
{ Fig.~\ref{fig:FDM-mH} shows the scatter plot of the allowed range for the electron EDM as a function of $m_{H_2}$, satisfying $1.5\times 10^{-9}\lesssim \Delta a_\mu\lesssim 4.0\times 10^{-9}$. Here the red and blue dots are for $0\le\phi\le\pi/4$ and $\pi/4< \phi\le\pi/2$, respectively. One finds that the electron EDM tends to grow as $m_{H_2}$ ($\phi$) increases (decreases), which directly follows from Eq.~(\ref{eq:edm}). 
 }

Our numerical studies show that $\text{BR}(\tau\to\mu\gamma)\lesssim 10^{-11}$,
which is two orders of magnitude smaller than the future sensitivity of $10^{-9}$ at Belle II~\cite{Aushev:2010bq}, 
while $\text{BR}(Z\to\mu\tau)$ lies just below the current experimental bound of $1.2\times 10^{-5}$ in Eq.~(\ref{eq:zmt-exp}) and larger than about $1.0\times10^{-7}$.
Therefore, the latter channel can be readily tested by a Giga-$Z$ type experiment at lepton colliders
(for earlier studies, see, {\it e.g.}, Ref.~\cite{AguilarSaavedra:2001rg}).
{ Notice here that the typical scales of $\text{BR}(Z\to e\mu)$ and $\text{BR}(Z\to e\tau)$ are $10^{-7}$, while $\text{BR}(\mu\to e\gamma)$ and $\text{BR}(\tau\to e\gamma)$ run over wide ranges, satisfying experimental upper bounds.
}

%%%%%%%%%%%%%%%%%%%
\begin{figure}[t]
\begin{center}
\includegraphics[width=100mm]{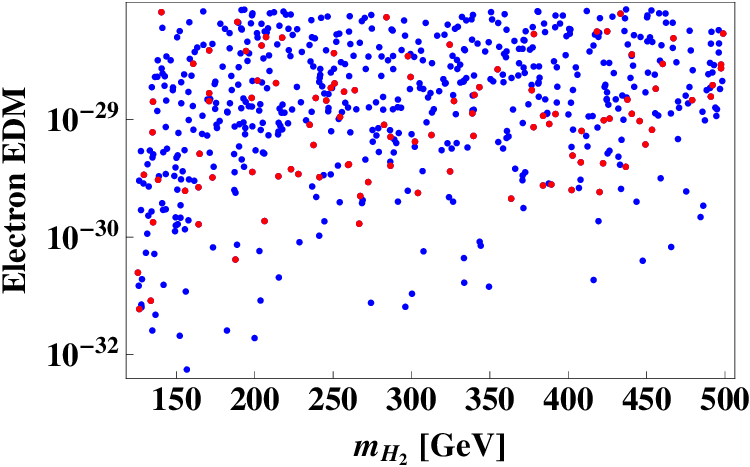} 
%%%
\caption{Scatter plot of the allowed range for the electron EDM as a function of $m_{H_2}$, satisfying $1.5\times 10^{-9}\lesssim \Delta a_\mu\lesssim 4.0\times 10^{-9}$. Here the red and blue dots are for $0\le\phi\le\pi/4$ and $\pi/4< \phi\le\pi/2$, respectively.
}
  \label{fig:FDM-mH}
\end{center}\end{figure}
%%%%%%%%%%%%%%%%%%%

}

%%%%%%%%%%%%%%%%%%%%%%%%%%%%%%%%%%%%%%%%%%%%%%%%%%
\section{Summary \label{sec:summary}}
%%%%%%%%%%%%%%%%%%%%%%%%%%%%%%%%%%%%%%%%%%%%%%%%%%

We have proposed a model of one-loop induced Majorana mass for neutrinos.  In analyzing the phenomenological aspects of the model, we have discussed radiative lepton decays with flavor violation, the muon anomalous magnetic moment, electric dipole moments (EDM's), Higgs to $\gamma\gamma$ decay, flavor-changing leptonic $Z$ decays, and scenarios with a bosonic or fermionc dark matter (DM) candidate.  We have scanned the parameter space to find experimentally allowed regions.  A nice feature of the model is that we can take an arbitrarily large scale for $M_N$ without affecting the neutrino oscillation data.  This enables the possibility of realizing baryogenesis via high-scale leptogenesis.

We conclude that one cannot get a sizeable contribution to the muon $g-2$ to match data in the bosonic DM scenario, since it conflicts with the constraints of both DM relic density and $\text{BR}(\mu\to e\gamma)$.  
In this case, the correction to muon $g-2$ is at most ${\cal O}(10^{-12})$, about three orders of magnitude smaller than the experimental bound.

For the fermionic DM scenario, on the other hand, we have shown that under various constraints it is possible to achieve $1.5\times 10^{-9}\lesssim \Delta a_\mu\lesssim 4.0\times 10^{-9}$ while satisfying the DM relic density and the direct detection bound provided that the DM mass is about $m_{H_1}/2$.  Remarkably, through parameter scanning we also have found that $\text{BR}(Z\to\mu\tau)$ often lies near the current experimental bound of $1.2\times 10^{-5}$, while $\text{BR}(\tau\to\mu\gamma)$ is well suppressed.
This is a testable smoking gun at future lepton colliders.

%%%%%%%%%%%%%%%%%%%%%%%%%%%%%%%%%%%
\begin{acknowledgments}
CWC would like to thank the hospitality of the Theoretical Particle Physics Group of Kyoto University during his visit when this work was finished.  CWC and ES were supported in part by the Ministry of Science and Technology (MOST) of R.O.C. under Grant Nos. MOST~104-2628-M-002-014-MY4 and MOST~104-2811-M-008-011, respectively.
\end{acknowledgments}

%%%%%%%%%%%%%%%%%%%%%%%%%%%%%%%%%%%

\end{document}